# Vector meson dominance and deep inelastic scattering at low and medium $Q^2$

Edgar V. Bugaev
*Institute for Nuclear Research of the Russian Academy of Sciences, Moscow 117312, Russia*

Boris V. Mangazeev
*Department of Physics, Irkutsk State University, Irkutsk 664003, Russia*

We argued that deep inelastic scattering (DIS) at small values of $Q^2$ is an essentially nonperturbative process and can be described, partially at least, by the vector meson dominance (VMD) model. We showed by the straightforward calculation that VMD model alone can successfully explain data on structure functions of DIS in a broad interval of $x$ ($5 \cdot 10^{-2} \div 10^{-4}$) for the region $Q^2 \leq 1\, GeV^2$. For a description of data at larger $Q^2$ we used the two-component (VMD + perturbative QCD) approach. We showed that these two components can be separated if VMD is used in the aligned jet version. We took into account, in calculations of VMD component of structure functions, the excited states of the $\rho$-meson and nondiagonal transitions between different members of the $\rho$-meson family. Amplitudes of these transitions were obtained using a formalism of the light-front Bethe-Salpeter equation and the method of diffraction-scattering eigenstates. The perturbative QCD component was calculated using a framework of the colour dipole model with the dipole cross section having a Regge-type energy dependence. We presented results of the detailed comparison of our predictions with experimental data for structure functions of the nucleon. We obtained also approximate predictions for the structure functions in the region of very small $x$, up to $10^{-8} \div 10^{-9}$, and showed that nonperturbative component at such values of $x$ is still relatively large and must be taken into account if $Q^2$ is about few $GeV^2$ or less.

## I. INTRODUCTION

It had been shown almost 50 years ago that the vector meson dominance hypothesis [1] can be successfully used for a description of photoproduction and low $q^2$ electroproduction processes. In particular, $\rho, \omega, \varphi$-dominance model applied to the forward Compton scattering amplitude connects the total photoproduction cross section, $\sigma_{\gamma p}$, with the total vector meson proton cross section, $\sigma_{Vp}$, ($V = \rho, \omega, \varphi$), and with diffractive vector meson photoproduction cross section, $d\sigma/dt(\gamma p \to Vp)$. This connection formulated as the "photoproduction sum rule" [2] had been checked by experiment and the agreement with data proved to be good enough, in a case of weakly virtual photons. However, strong discrepancies with experiment had been observed in a case of large spacelike $q^2$, and this had been interpreted as a reveal of the coupling of the photon to higher mass states. Naturally, the heavy masses become relatively important with an increase of $q^2$, due to propagator factors, $1/(Q^2 + M^2)^2$, in VMD sums (for spacelike photons $Q^2 = -q^2 > 0$ in our metric). The first heavy vector mesons had been observed in $e^+e^-$-annihilation experiments around 1972, and, at the same time, the approach named as "Generalized Vector Dominance" (GVD) had appeared [3,4]. The GVD models use, to describe the structure functions $\sigma_{T,L}$ of the inelastic electron scattering, spectral representations for imaginary parts of transverse and longitudinal forward Compton amplitudes, the corresponding spectral weight functions, $\rho_{T,L}$, being related with the amplitudes for $Vp \to V'p$ scattering:

$$\sigma_{T,L}(Q^2, s) = \iint dM^2 dM'^2 \frac{\rho_{T,L}(s, M^2, M'^2) M^2 M'^2}{(M^2 + Q^2)(M'^2 + Q^2)}. \quad (1)$$

This form implies that, in principle, in nature there can be even infinitely many mesons (like in $N_c \to \infty$ limit of QCD [5]). According to VMD concept these mesons are the same as the mesons produced in the process $e^+ e^- \to hadrons$. The hypothesis of the quark-hadron duality [6] suggests that the observed scaling in $e^+e^- \to hadrons$, i.e., the behavior $\sigma(e^+e^- \to hadrons) \sim 1/s$, is just a consequence of the infinitely large number of vector mesons. Surely, very heavy vector mesons have large hadronic widths and, being produced in $e^+e^-$-annihilation, they merge in the hadronic continuum. But in some cases one can consider vector mesons as narrow or even zero-width resonances. For example, in VDM models the sums over meson masses converge rather well as we will see below and in this case the zero-width approximation is justified.

VMD, combined with the quark-hadron duality, is used, e.g., also for a description of two-point functions (vector current-vector current correlators) and electromagnetic pion formfactor (see, e.g., [7]). Matching of the VMD predictions with the corresponding pQCD and OPE formulas at large $Q^2$ leads, again, to a requirement of infinitely many vector mesons. In all such calculations, beginning from the pioneering work [4], the mass spectrum of vector mesons was taken in a simple form:

$$M_n^2 = M_0^2 (1+an). \qquad (2)$$

Such a form arises, e.g., in QCD string models, as a result of semi-classical quantization of a straight string system (see [8] and references therein). The same spectrum (with a=2) had been predicted by the Veneziano model [9] and it is often referred to as a" radial Regge mass spectrum". Just this form of the vector meson mass spectrum, as had been shown in [4, 6, 10], is needed to reproduce rather well the partonic logarithm of the two-point correlator.

The dominance of $\rho, \omega, \varphi$-mesons in electromagnetic interactions of hadrons at low energies has a very solid theoretical explanation. Two main phenomena are in need of such an explanation: i) the direct $\gamma hh$-coupling is absent at lowest order in the hadron momentum, the entire photon coupling being released through a virtual vector meson, and ii) the vector mesons are coupled to conserved hadronic currents. It had been shown in many works of last century that both these features are predicted by field theories operating with effective chiral Lagrangian of pseudoscalars and vector mesons (see, e.g., [11] for a review). In particular, all predictions of VMD were reproduced in the model [12], in which $\rho$-meson arises as the dynamical gauge boson of a hidden local symmetry (HLS) in the nonlinear chiral Lagrangian, and the mass of the $\rho$ is generated by spontaneous breaking of the chiral symmetry through the Higgs mechanism. There are other field-theory models of vector mesons which are motivated by the VMD, e.g., the massive Yang-Mills models [13] and the WCCWZ-approach [14] using the idea of nonlinear realization of chiral symmetry.

It is remarkable that in HLS model [12] $\rho$ meson is a gauge boson as in the conjecture of Sakurai suggested in 1960 [15]. Nowadays it became clear, as we tried to argue in this Section, that one needs field-theory models with an infinite number of vector mesons. Fortunately, in last ten years the new theoretical approach to modeling low energy properties of QCD had appeared consisting in studies of five-dimensional holographic duals of QCD [16, 17]. The idea (which goes back to the work [18]) is to reproduce holographically the important properties of QCD, such as a confinement and chiral symmetry breaking. One of the consequences of such a description is just the VMD, in which the towers of vector mesons including all excited states, i.e., $\rho, \omega, \varphi$ with their families, contribute. It is essential that the vector mesons of these towers are *gauge* fields with *hidden gauge* invariance, i.e., the five-dimensional approach of [16, 17] is the natural generalization of earlier ideas of works [15, 12] suggested to explain the $\rho, \omega, \varphi$-dominance. The works [16, 17] not only marked the "return of vector meson dominance" [19], these works propose the field theory basis for GVD predicted in [3] more than 40 years ago.

The main aim of the present paper is a calculation of the VMD contribution in structure functions of DIS. Traditionally, VMD had been used for predictions of structure function of DIS in the diffraction region [20, 21] of the $v - Q^2$ plane, i.e., in the region of small $x$, $x \square 0.1$ (irrespectively of the $Q^2$-value). This limitation is connected with the necessary condition

$$2v/(Q^2 + m^2) \square 2R,$$

where $m$ is the characteristic mass value of the hadron in the photon's fluctuation, $R$ is the radius of the target.

It is very important to understand that the mass spectrum of photon fluctuations can never be saturated by the vector mesons alone. Quark-antiquark pair interacts with the target as a vector meson if the transverse momentum $k_\perp$ of the pair's quark is not large and confinement effects are essential. If $k_\perp$ is small, the pair is similar with a jet aligned along the photon's momentum [22], but the transverse size of it becomes, due to the evolution, of order of the hadronic size, before an arrival at the target. Only such pairs can be considered as "vector mesons", the quotation marks here signify that these mesons, in contrast with a case of the usual hadrons, interact solely nonperturbatively with the target. This conclusion is consistent with the fact that VMD is an essentially low energy approach and cannot be applied to a description of hard processes such as processes of jet production in meson-nucleon collisions. Evidently, perturbative contributions to the structure functions must be calculated separately using some model. So, the whole approach is necessarily the two-component one.

In literature there are examples of one-component models describing the structure functions of DIS without exploiting the VMD. The model GVD-CDP (Generalized Vector Dominance – Colour Dipole Picture) [23] uses the framework of the colour dipole model [24] with a QCD-inspired ansatz for the $(q\bar{q})p$ forward scattering amplitude. The model imitates destructive interference effects which usually must be incorporated in off-diagonal vector dominance models [25] to provide convergence of sums over vector meson states. Note that in spite of the name, the model is non-hadronic: it operates with quarks and gluons only.

Another way of the one-component description of DIS is the approach based on pQCD. In particular, it had been realized (see, e.g., [26]) that non-perturbative effects in DIS are partly masked by effects of gluon saturation if the saturation scale is relatively large (for the lowest $Q^2$ data at HERA $Q_s^2 \sim 2\ GeV^2$ for $x \sim 10^{-5}-10^{-6}$). Nevertheless, the basic fact is that the colour dipole-proton scattering amplitude is a genuinely non-perturbative object and it is impossible, performing a "global fit" to the structure functions, to avoid a modeling and a use of phenomenological ansatzes [27].

At the end of this introduction one should mention several works where the separation of soft and hard components in structure functions of DIS have been performed in a way which is closest to ours. Authors of [28,29] use a Regge-type energy dependence for both components of $\sigma_{\gamma p}$, and the corresponding intercepts are different ($\alpha_P \sim 1.06-1.08$ for the soft component and $\alpha_P \sim 1.3-1.4$ for the hard one). The paper [28] uses for both components the formalism of GVD (eq. (1)), in diagonal approximation, whereas papers [29] use the colour dipole model. Authors [30,31] use for a description of the soft component the VMD in its simplest form (only $\rho, \omega, \varphi$-mesons are taken into account). Their criterion of the separation differs from ours: it is assumed that $M_{q\bar{q}}^2$ is a good measure of the transverse size for a majority of $q\bar{q}$-pairs and AJM-like contributions are small.

Preliminary results of calculations with the two-component approach developed in the present paper have been published in works [32].

The plan of the paper is as follows. In the second Section the main formulas of non-diagonal VMD model, in the aligned jet version, are obtained, starting from the colour dipole model and using the quark-hadron duality arguments. In the third Section the formalism of the light-front Bethe-Salpeter equation is used for an obtaining the approximate expressions for the vector meson mass spectrum and the meson's wave functions. Further, using a method of the diffraction scattering eigenstates the expressions for the amplitudes of non-diagonal transitions ($Vp \to V'p$) are derived. In the fourth Section the structure functions of DIS are calculated using, for a soft component, the non-diagonal VMD and, for a hard component, the Regge-type parameterization suggested in [29]. The last Section contains our conclusions.

## II. VMD IN ALIGNED-JET VERSION

The starting point of our consideration is the expression based on the perturbative QCD and two-step picture of the $\gamma^* p$ interaction: the $\gamma^* \to q\bar{q}$ conversion is followed by an interaction of the $q\bar{q}$-pair with the target proton (Fig. 1). For definiteness we use here the GVD-CDP model [23].

The total $\gamma^* p$ interaction cross section (summed over all possible final hadronic states) is given by the contribution of the $q\bar{q}$-channel in the imaginary part of the Compton forward scattering amplitude,

$$\sigma_{T,L}(Q^2,s) = \frac{1}{16\pi} \sum_j \sum_{r,r'} \int dz \int d^2k_\perp \int dz' \int d^2k'_\perp$$
$$\times \psi_{\gamma(r,r')}^{T,L*}(\vec{k}'_\perp, z', Q^2) \qquad (3)$$
$$\times \frac{1}{s} A_{q\bar{q}\to p}(\vec{k}'_\perp, z', \vec{k}_\perp, z, s) \psi_{\gamma(r,r')}^{T,L}(\vec{k}_\perp, z, Q^2).$$

In this expression $A_{q\bar{q}\to p}$ is an imaginary part of the $(q\bar{q})p$ forward scattering amplitude, $Q^2$ and $\sqrt{s}$ are a virtuality of the photon and $\gamma^* p$ center-of-mass energy, $\psi_{\gamma(r,r')}^{T,L}$ are light cone wave functions of $q\bar{q}$-fluctuations of the virtual photon with transverse or longitudinal polarization. These wave functions depend on quark and antiquark helicities $(r/2, r'/2)$, quark mass ($m_q$) and quark momentum variables $z, k_\perp$ ($z = k_+/q_+$ is the fraction of the photon light cone momentum carried by the quark, $k_\perp$ is the transverse momentum of the incoming quark).

Using $M_{q\bar{q}}$ and $z$ as independent variables rather than $k_\perp, z$, where $M_{q\bar{q}}$ is an invariant mass of the $q\bar{q}$-pair,

$$M_{q\bar{q}}^2 = \frac{k_\perp^2 + m_q^2}{z(1-z)}, \qquad (4)$$

one can show that the wave functions contain the transverse and longitudinal electromagnetic currents $j_T^{r,r'}$ and $j_L^{r,r'}$ and quark propagators,

$$\psi_{\gamma(r,r')}^{T,L} \sim e_q \frac{j_{T,L}^{r,r'}}{Q^2 + M_{q\bar{q}}^2}, \qquad (5)$$

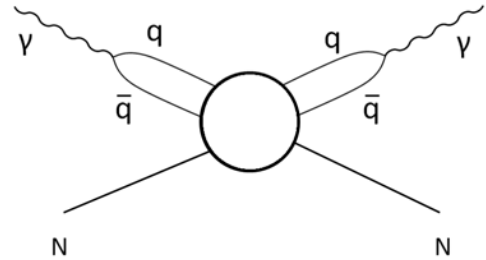

FIG. 1. The diagram representing schematically the contribution of $q\bar{q}$-channel in an imaginary part of the Compton forward scattering amplitude.

and, in turn, these currents are functions of $z$ [33],
$$j_T^{r,r'} \Box (2z+1\pm r)\delta_{r,r'}, \quad j_L^{r,r'} \Box \sqrt{z(z-1)}\,\delta_{r,r'}. \qquad (6)$$

The final expressions for $\sigma_{T,L}$ depend on a square of $e_q j_{T,L}$, averaged on $r, r'$,

$$e_q^2 \sum_{r,r'} \left| j_T^{r,r'} \right|^2 \Box\, e_q^2 \left[ z^2 + (1-z)^2 \right],$$
$$e_q^2 \sum_{r,r'} \left| j_L^{r,r'} \right|^2 \Box\, e_q^2 z(1-z). \tag{7}$$

The main assumption of the GVD-CDP approach is that the amplitude $\frac{1}{s} A_{q\bar{q} \to p}$ is similar in structure with corresponding amplitude predicted by the two-gluon-exchange approximation [34, 35]. Namely, the simple expression:

$$\frac{1}{s} A_{q\bar{q} \to p}\left(\vec{k}_\perp', z', \vec{k}_\perp, z, s\right) = 2(2\pi)^3 \int d^2 l_\perp \tilde{\sigma}_{q\bar{q}p}\left(l_\perp^2, z, s\right)$$
$$\times \left[ \delta\!\left(\vec{k}_\perp' - \vec{k}_\perp\right) - \delta\!\left(\vec{k}_\perp' - \vec{k}_\perp - \vec{l}_\perp\right) \right] \delta(z - z') \tag{8}$$

is suggested. Here, $\tilde{\sigma}_{q\bar{q}p}$ is the color dipole cross section, $|\vec{l}_\perp|$ is the transverse momentum transfer. The simplest and very convenient ansatz for the colour dipole cross section is [36]

$$\tilde{\sigma}_{q\bar{q}p}\left(l_\perp^2, z, s\right) = \sigma_0(s) \delta\!\left(l_\perp^2 - z(1-z)\Lambda^2\right). \tag{9}$$

We assume that $\Lambda$ is an energy independent constant (i.e., it doesn't depend on $s$). It will be clear below that this assumption is necessary if one wants to have the duality of GVD-CDP with our formulation of VMD.

The position space colour dipole cross section is given by the formula

$$\sigma_{q\bar{q}p}\left(r_\perp^2, z, s\right) = \sigma_0(s)\left(1 - J_0\!\left(r_\perp \sqrt{z(1-z)}\Lambda\right)\right). \tag{10}$$

It is proportional to $r_\perp^2$ at small $r_\perp^2$ (colour transparency) and goes to a constant at $r_\perp^2 \to \infty$ (it gives hadronic unitarity if the $s$-dependence of $\sigma_0(s)$ is not too strong).

Ansatz (9), together with eqs. (7), leads to very simple final expressions for $\sigma_{T,L}(Q^2, s)$ (everywhere below we simplify the notation, $M_{q\bar{q}} \to M$):

$$\sigma_T\!\left(Q^2, s\right) \cong \frac{e^2}{12\pi^2} R_{e^+e^-} \sigma_0(s) \int_0^1 dz\, \frac{3}{2}\left[z^2 + (1-z^2)\right]$$
$$\times \left\{ \int dM^2 \frac{M^2}{\left(Q^2 + M^2\right)^2} + nondiagonal\ part \right\}, \tag{11}$$

$$\sigma_L\!\left(Q^2, s\right) \cong \frac{e^2}{12\pi^2} R_{e^+e^-} \sigma_0(s) \int_0^1 dz\, 6z(1-z)$$
$$\times \left\{ \int dM^2 \frac{Q^2}{\left(Q^2 + M^2\right)^2} + nondiagonal\ part \right\}. \tag{12}$$

Here,

$$R_{e^+e^-} = N_c \sum_q \left(\frac{e_q}{e}\right)^2. \tag{13}$$

Nondiagonal parts in eqs. (11, 12) depend on $M^2$, $M'^2$, $\Lambda$. The integrals over $z$ in these equations are equal to 1 and we omit these factors temporarily. We see that the $\gamma q\bar{q}$-coupling is completely determined by the quark's charges and the pair's mass.

For connection of this approach with VMD one must introduce the vector mesons, the $\gamma V$-coupling and vector-meson nucleon amplitudes, using quark-hadron duality arguments. For simplicity, we consider only one vector meson family: $\rho$-meson and its excitations. We assume that the mass spectrum of the $\rho$-family is equidistant in a square of mass, i.e.,

$$M_n^2 = M_\rho^2 (1 + an),$$
$$\Delta M^2 = M_{n+1}^2 - M_n^2 = M_\rho^2 a. \tag{14}$$

The integral in eq. (11) can be rewritten in a form

$$\int \frac{dM^2}{M^2} \frac{M^4}{\left(Q^2 + M^2\right)^2} \cong \sum_i \frac{\Delta M^2 M_i^4}{M_i^2 \left(Q^2 + M_i^2\right)^2}. \tag{15}$$

Using this form one obtains for the diagonal part of $\sigma_T$:

$$\sigma_T\!\left(Q^2, s\right) = \frac{e^2}{12\pi^2} R_{e^+e^-} \sum_i \frac{\Delta M^2 M_i^4}{M_i^2 \left(Q^2 + M_i^2\right)^2} \sigma_0(s)$$
$$\cong \sum_i \frac{e^2}{f_i^2} \frac{M_i^4}{\left(Q^2 + M_i^2\right)^2} \sigma_0(s), \tag{16}$$

where

$$\frac{e^2}{f_i^2} \equiv \frac{e^2}{12\pi^2} R_{e^+e^-} \frac{\Delta M^2}{M_i^2}. \tag{17}$$

Now, the quark-hadron duality suggests the replacement of a sum over $q\bar{q}$-pairs with a mass $M_i$ in eq. (16) by a sum over vector mesons with a mass $M_{V_n} \equiv M_n$ and, correspondingly, an introduction of $\gamma V_n$-coupling constants $\frac{e}{f_{V_n}} \equiv \frac{e}{f_n}$ defined by the expression [37]

$$\frac{e^2}{f_n^2} \equiv \frac{e^2}{12\pi^2} R_{e^+e^-} \frac{\Delta M^2}{M_n^2}, \tag{18}$$

where $\Delta M^2$ is determined now from the vector meson mass spectrum, eq. (14).

Finally, one obtains the familiar GVD expressions for $\sigma_{T,L}(s)$ (in diagonal approximation):

$$\sigma_T\!\left(Q^2, s\right) = \sum_n \frac{e^2}{f_n^2} \frac{M_n^4}{\left(Q^2 + M_n^2\right)^2} \sigma_n^T(s), \tag{19}$$

$$\sigma_L\!\left(Q^2, s\right) = \sum_n \frac{e^2}{f_n^2} \frac{Q^2 M_n^2}{\left(Q^2 + M_n^2\right)^2} \sigma_n^L(s). \tag{20}$$

Here, $\sigma_n^{T,L}$ are total cross sections for $V_n$-nucleon interactions for the vector mesons with transverse (T) and longitudinal (L) polarizations.

Now we should take into account the fact that the quark-hadron duality approach used in a derivation of eqs. (19, 20) is based on the VMD. It means, in particular, that cross sections $\sigma_n(s)$ correspond to nonperturbative processes only. However, the $q\bar{q}$-pair with a small transverse size being colour neutral weakly interacts with a hadronic target, this interaction is calculated in the framework of perturbative QCD. According to the uncertainty principle

$$r_\perp^2 \Box \frac{1}{k_\perp^2} \Box \frac{1}{M^2 z(1-z)} \qquad (21)$$

(for massless quarks). So, at fixed and not very small values of $M^2$, quarks of the wide $q\bar{q}$-pairs (those having large $r_\perp^2$ and interacting nonperturbatively) have relatively small transverse momenta and are asymmetric in the longitudinal energy, that is $z(1-z)$ is relatively small.

To separate approximately perturbative and nonperturbative interactions of the $q\bar{q}$-pair with the nucleon target we introduce the model parameter $k_{0\perp}^2$. We assume that $q\bar{q}$-pair interacts nonperturbatively if transverse momentum of pair's quarks is smaller than $k_{0\perp}$. For a given pair's mass $M$ it means that

$$z(1-z) < \frac{k_{0\perp}^2}{M^2}. \qquad (22)$$

This criterion constrains the variable $z$, so, now one must return to integrals over $z$ in eqs. (11, 12). The constraint (22) leads to the following changes (if we want to keep in the structure functions $\sigma_{T,L}$ only nonperturbative parts):

$$\frac{3}{2}\int_0^1 dz \left[z^2 + (1-z)^2\right] = 1 \to \frac{3}{2}\int_0^1 dz \left[z^2 + (1-z)^2\right]$$
$$\times \Theta\left(\frac{k_{0\perp}^2}{M^2} - z(1-z)\right) = \eta_T\left(M^2, k_{0\perp}^2\right), \qquad (23)$$

$$6\int_0^1 z(1-z)dz = 1 \to 6\int_0^1 z(1-z)$$
$$\times \Theta\left(\frac{k_{0\perp}^2}{M^2} - z(1-z)\right)dz = \eta_L\left(M^2, k_{0\perp}^2\right). \qquad (24)$$

The straightforward calculation gives the following values of $\eta$-factors

$$\eta_T\left(M^2, k_{0\perp}^2\right) = 3z_0 - 3z_0^2 + 2z_0^3, \qquad (25)$$

$$\eta_L\left(M^2, k_{0\perp}^2\right) = 6z_0^2 - 4z_0^3, \qquad (26)$$

$$z_0 = \frac{1}{2} - \sqrt{\frac{1}{4} - \frac{k_{0\perp}^2}{M^2}} \Box \frac{k_{0\perp}^2}{M^2}. \qquad (27)$$

Evidently, the same factors (we will call them "cut-off factors") appear also in VMD expressions for $\sigma_{T,L}(Q^2, s)$:

$$\sigma_T(Q^2, s) = \sum_n \frac{e^2}{f_n^2} \eta_T\left(M_n^2, k_{0\perp}^2\right) \frac{M_n^4}{\left(Q^2 + M_n^2\right)^2} \sigma_n^T(s), \qquad (28)$$

$$\sigma_L(Q^2, s) = \sum_n \frac{e^2}{f_n^2} \eta_L\left(M_n^2, k_{0\perp}^2\right) \frac{Q^2 M_n^2}{\left(Q^2 + M_n^2\right)^2} \sigma_n^L(s). \qquad (29)$$

Physically, one can say that the cut-off factors lead to a strong decreasing of the $\gamma V$-coupling for heavy vector mesons in DIS processes because only the $q\bar{q}$-pairs with large transverse size are taken into account. Really, $\eta_{T,L} \Box 1$, so

$$\frac{e^2}{f_n^2} \eta_{T,L} \equiv \frac{e^2}{\left(f_n^{*2}\right)_{T,L}} \Box \frac{e^2}{f_n^2}. \qquad (30)$$

The duality relation (18) is modified now, in application of VMD to DIS, to

$$\frac{e^2}{\left(f_n^{*2}\right)_{T,L}} = \eta_{T,L} \frac{e^2}{12\pi^2} R_{e^+e^-} \frac{\Delta M^2}{M_n^2}. \qquad (31)$$

If we consider, in VMD approach, an interaction of the $q\bar{q}$-pair with the nucleon as an interaction of the vector meson, then, surely, VMD formulas for the structure functions $\sigma_{T,L}$ must contain also the nondiagonal contributions (Fig. 2). In GVD-CDP picture the nondiagonal transitions of $q\bar{q}$-pairs are quite essential leading to large cancellations in final formulas. In contrast with this, it is well known by the experience of hadron physics that in our VMD such cancellations cannot be too large. We study a role of nondiagonal transitions, $V_n p \to V_{n'} p$, in Section 4. The general formulas for $\sigma_{T,L}$ containing the nondiagonal contributions are:

$$\sigma_T(Q^2, s) = \sum_n \frac{e^2}{f_n^2} \eta_T\left(M_n^2, k_{0\perp}^2\right) \frac{M_n^4}{\left(Q^2 + M_n^2\right)^2} \sigma_n^T(s)$$
$$+ \sum_{n\neq n'} \frac{e^2}{f_n f_{n'}} \eta_T\left(\max\left(M_n^2, M_{n'}^2\right), k_{0\perp}^2\right) \qquad (32)$$
$$\times \frac{M_n^2 M_{n'}^2}{\left(Q^2 + M_n^2\right)\left(Q^2 + M_{n'}^2\right)} \frac{1}{s} \operatorname{Im} F_{n,n'}^T(s),$$

$$\sigma_L(Q^2, s) = \sum_n \frac{e^2}{f_n^2} \eta_L\left(M_n^2, k_{0\perp}^2\right) \frac{Q^2 M_n^2}{\left(Q^2 + M_n^2\right)^2} \sigma_n^L(s)$$
$$+ \sum_{n\neq n'} \frac{e^2}{f_n f_{n'}} \eta_L\left(\max\left(M_n^2, M_{n'}^2\right), k_{0\perp}^2\right) \qquad (33)$$
$$\times \frac{M_n^2 M_{n'}^2}{\left(Q^2 + M_n^2\right)\left(Q^2 + M_{n'}^2\right)} \frac{1}{s} \operatorname{Im} F_{n,n'}^L(s).$$

Here, we introduce the notation $F_{n,n'}^{T,L}$ for amplitudes of $V_n p \to V_{n'} p$ scattering for mesons with transverse and longitudinal polarizations.

It is easy to prove that, due to a presence of the cut-off factors in sums over vector mesons in eqs. (32, 33) the convergence takes place even if no cancellations arise after addition of nondiagonal terms.

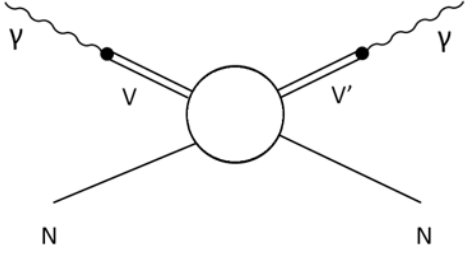

FIG. 2. The schematic diagram for the Compton forward scattering amplitude in a nondiagonal VMD model.

## III. THE HADRONIC AMPLITUDES AND WAVE FUNCTIONS

### A. Reduced Bethe-Salpeter equation

Vector mesons are bound states of two constituent quarks and for a description of their interactions with nucleons and for calculations of the mass spectra of their excited states it is quite convenient to use a formalism of the light front Bethe-Salpeter (BS) equation. The initial four-dimensional BS equation has a view [38]

$$i(2\pi)^4 \Phi(P,q) = \frac{1}{\Delta_1 \Delta_2} \int d^4 q' K(q-q') \Phi(P,q'). \quad (34)$$

In this equation $\Phi(P,q)$ is the BS wave function, $P$ and $q$ are total and relative momenta of meson's quarks, respectively,

$$P = p_1 + p_2, \quad q = \frac{1}{2}(p_1 - p_2). \quad (35)$$

$p_{1,2}$ are quark's momenta, $\Delta_{1,2} = p_{1,2}^2 - m_q^2$. We work in the approximation of free propagators and spinless meson and quarks.

Light front dynamics [39] (for a review see, e.g., [40]) operates with three inner momentum variables (we use the on-mass-shell condition, $P^2 = M^2$ (in this Section $M$ is a mass of the two-quark bound state), and neglect the transverse momentum of the meson as a whole, i.e., we put $P_\perp = 0$):

$$\vec{q}_\perp = \frac{1}{2}(\vec{p}_{1\perp} - \vec{p}_{2\perp}) = \vec{p}_{1\perp} = -\vec{p}_{2\perp}; \quad y = \frac{q_+}{P_+}. \quad (36)$$

These three internal variables essentially represent the relative momenta, $\vec{q}_\perp, q_+$, of two constituent quarks. For the variable $y$ one has from eqs. (35, 36):

$$y = \frac{p_{1+} - p_{2+}}{2 P_+}. \quad (37)$$

The light front reduction of the BS equation consists in integrating both sides of eq. (34) over $q_-$. We assume that the interaction kernel doesn't depend on $q_-$ ("instantaneous approximation"), i.e.,

$$K(q-q') \Box K(\vec{q}_\perp - \vec{q}_\perp', y - y'). \quad (38)$$

We define the reduced BS wave function by the relation [41]

$$\psi(q_\perp, y) = \int \frac{1}{2} dq_- \Phi(P,q). \quad (39)$$

Integration of the product of two propagators standing in the right hand side of the BS equation (34) gives the well-known result (see, e.g., [42]).

$$\int \frac{1}{2} dq_- (\Delta_1 \Delta_2)^{-1} = \frac{2\pi i}{2 P_+} \left(-q_\perp^2 - m_q^2 + M^2 x(1-x)\right)^{-1}. \quad (40)$$

Here, the variable $x$ is introduced,

$$x = \frac{1}{2} + y = \frac{p_{1+}}{P_+}, \quad 0 < x < 1. \quad (41)$$

Using the relation

$$d^4 q = \frac{1}{2} dq_- dq_+ dq_\perp \quad (42)$$

and the definition (39) one obtains finally the reduced BS equation :

$$\left[M^2 x(1-x) - \left(q_\perp^2 + m_q^2\right)\right]\psi(\vec{q}_\perp, y)$$
$$= \frac{1}{2(2\pi)^3} \int d^2 q_\perp' dy' K(\vec{q}_\perp - \vec{q}_\perp', y - y')\psi(\vec{q}_\perp', y') \quad (43)$$

This equation is equivalent to the corresponding equation obtained in a framework of the effective light cone QCD-inspired theory [43].

### B. Spectrum of vector meson excitations

The next step is an introduction of the approximate confinement scheme. The simplest phenomenological Lorentz invariant model of confinement had been suggested in [44]. If one units the three variables $\vec{q}_\perp, y$ in a single three vector

$$\vec{q} = (q_{1\perp}, q_{2\perp}, \mu y) \quad (44)$$

($\mu$ is a model parameter having a dimension of mass) one can introduce the angular momentum operator

$$\vec{J} = -i\vec{q} \times \vec{\nabla}_q. \quad (45)$$

After this one takes the three-dimensional potential which becomes infinite at large $|\vec{q}|$ to secure that eigenstates of the mass operator are confined to a region $\vec{q}^2 < \vec{q}_c^2$.

Using (44), the equation (43) can be rewritten in the form

$$(2\pi)^3 \left[-\frac{M^2}{4} + m_q^2 + \vec{q}^2\right]\psi(\vec{q})$$
$$= \frac{1}{2\mu} \int d^3 q' K(\vec{q} - \vec{q}')\psi(\vec{q}'). \quad (46)$$

We suppose that the kernel $K(\vec{q} - \vec{q}')$ describes the long range quark-antiquark interaction of oscillatory type leading to the confinement (see, e.g., [41] and references therein),

$$K(\vec{q}-\vec{q}\,') = (2\pi)^3 \, \omega_{q\bar{q}}^2 \left( \vec{\nabla}_q^2 + \omega_0^{-2} \right) \delta^3(\vec{q}-\vec{q}\,'). \quad (47)$$

In $\vec{r}$-space one has, correspondingly,

$$K(\vec{r}) = \frac{1}{(2\pi)^3} \int e^{-i\vec{q}\vec{r}} K(\vec{q}) d^3q = -\omega_{q\bar{q}}^2 \delta(\vec{r} - \omega_0^{-2}). \quad (48)$$

Two parameters of this kernel, $\omega_{q\bar{q}}^2$ and $\omega_0^{-2}$, (the latter gives the shift of the potential at $\vec{r}=0$) can be determined from data for a mass spectrum of bound states.

Substituting the kernel $K(\vec{q}-\vec{q}\,')$ from eq. (47) in eq. (46) one obtains the equation

$$\left( -\frac{M^2}{4} + m_q^2 + \vec{q}^{\,2} \right) \psi(\vec{q}) = \frac{\omega_{q\bar{q}}^2}{2\mu} \left( \vec{\nabla}_q^2 + \omega_0^{-2} \right) \psi(\vec{q}). \quad (49)$$

Now it is convenient to use the new variable $\beta$ (with the dimension of energy) defined by the relation

$$\frac{\omega_{q\bar{q}}^2}{2\mu} \equiv \beta^4. \quad (50)$$

With this new variable the eq. (49) is rewritten in the form

$$\left( -\frac{\beta^2}{2} \vec{\nabla}_q^2 + \frac{q^2}{2\beta^2} \right) \psi(\vec{q})$$
$$= \frac{1}{2\beta^2} \left( \frac{M^2}{4} - m_q^2 + \beta^4 \omega_0^{-2} \right) \psi(\vec{q}). \quad (51)$$

The full factor in front of $\psi(\vec{q})$ in the right hand side of this equation is dimensionless constant, so, formally it is the equation for a wave function of a particle moving in the three-dimension oscillatory potential. Eigenstates of mass operator square are obtained from the quantum condition

$$\frac{1}{2\beta^2} \left( \frac{M^2}{4} - m_q^2 + \beta^4 \omega_0^{-2} \right) = N + \frac{3}{2}, \quad N = 2n + l, \quad (52)$$

where $n, l$ are radial and orbital quantum numbers respectively.

For simplicity we consider all excitations as radial ones, and put $l = 0$ everywhere below. The most essential feature of the meson mass spectrum in the present rough model is the equidistance in a square of mass:

$$M_{\rho_n}^2 = a(1 + bn), \quad (53)$$

$a = M_{\rho_0}^2$, $n = 0, 1, 2, \ldots$, $b \cdot M_{\rho_0}^2 = M_{\rho_n}^2 - M_{\rho_{n-1}}^2 = const.$

As is pointed out in the Introduction, such a behavior is predicted, in particular, by QCD string models. According to [45], the experimental values of rho-family masses are following (from $n = 0$ up to $n = 4$, in $MeV$): 770, 1450, 1700, 1900 and 2150. It is remarkable that, if we parameterize the mass spectrum by the formula (53), then, from a comparison with these experimental mass values, we obtain, using a least-square method, the value 1.76 for the parameter b, which is close to the value b=2 predicted, originally, by the Veneziano model [9] and used in many works exploiting the quark-hadron duality hypothesis (see the Introduction).

For calculations in VMD models it is better to use the mass spectrum parameters which give more accurate mass values for the lowest excited states (because just these states are most essential in VMD sums). Having this in mind, the approximate values of model parameters $\beta$, $\omega_0$ can be determined substituting in eq. (52) experimental values of meson masses for $n = 0$ and $n = 1$ (or for $n = 0$ and $n = 2$). It gives, in the first case, the following system of equations:

$$\frac{M_{\rho_0}^2}{4} - m_q^2 + \beta^4 \omega_0^{-2} = \frac{3}{2} 2\beta^2, \quad (54)$$

$$\frac{M_{\rho_1}^2}{4} - m_q^2 + \beta^4 \omega_0^{-2} = \left(2 + \frac{3}{2}\right) 2\beta^2.$$

If $m_q = 0.3 \, GeV$ one finds from here that $b = 2.5$ and $\beta^2 = 0.094 \, GeV^2$, $\omega_0^2 = 0.04 \, GeV^2$. In the second case, one has $b = 2$, $\beta^2 = 0.074 \, GeV^2$, $\omega_0^2 = 0.033 \, GeV^2$.

In fig.3 we show the experimental values of masses for numbers of rho-family, together with the lines for $b = 2$ and for $b = 2.5$.

Note that in eq.(52) the quark mass term enters in a combination with the parameter $\omega_0$, so, any change of the $m_q$ value is connected with a corresponding change of a value of $\omega_0$, so, a sensitivity of the mass spectrum to a value of the quark mass is weak. Note also that in constituent quark models of hadrons (see, e.g., [43]) the typically used values of the constituent quark mass are within a rather narrow interval, (0.26 - 0.31) $GeV$.

For concrete calculations we choose the following values of $\beta$ and $b$:

$$\beta^2 = 0.094, \quad b = 2.$$

Using such a choice we try to combine two cases, more or less. Although, our experience shows that final results very weakly depend on the $\beta^2$ value if it changes within the interval (0,074-0.094) $GeV^2$.

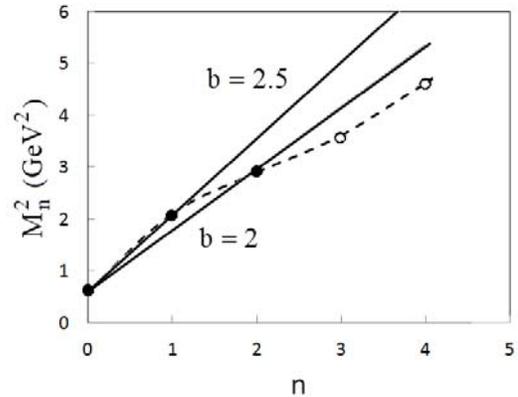

FIG. 3. The dependence of $M^2_{\rho_n}$ on the radial quantum number $n$ of the excited state. The dots are experimental values of vector meson masses of the $\rho$-family members (the values of masses for $n=3$ and for $n=4$ are poorly known). Lines correspond to two different parameterizations of mass spectrum (eq. (53)).

### C. Wave functions of vector mesons

The eigenfunctions of eq. (51) are well known (see, e.g., [46]). They are proportional to exponential factor $e^{-q^2/2\beta^2}$ (in momentum space) and to polynomial in $q = |\vec{q}|$. Radial part of the total wave function (we keep for it the same notation, $\psi$) depends on the modulus of $\vec{q}$,

$$\psi_n(q) = \frac{1}{\sqrt{A_n}} \sum_{k=0}^{n} \frac{(-1)^k}{4^k k!(2n+1-2k)!} \left(\frac{q}{\beta}\right)^{2n-2k} e^{-\frac{q^2}{2\beta^2}}. \quad (55)$$

Here $A_n$ is the normalization coefficient.

Below it will be convenient to separate in (55) $q_\perp$- and $y$-variables using the connection

$$q^2 = q_\perp^2 + \mu^2 y^2, \quad (56)$$

and after this to return to the initial notation, $\psi(q_\perp, y)$.

The numerical value of the model parameter $\mu$, introduced above, in eq, (44) is not very essential for final results. We suppose that $\mu$ is approximately equal to the mass of the ground state of the family.

The simplest way for a normalization of the reduced BS function is to calculate the electromagnetic formfactor $F(Q^2)$ of the bound state in this light front formalism and to put $F(0) = 1$. The corresponding diagram is shown in Fig. 4. It contains two vertex functions and three quark propagators.

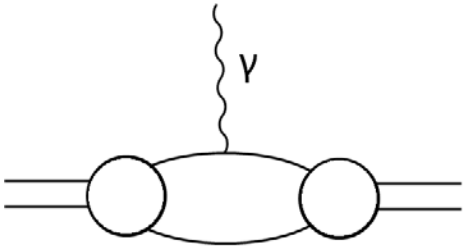

FIG. 4. The diagram used for a normalization of the BS wave function.

The four-dimensional vertex function is defined by the relation

$$\Phi(P,q) = \frac{1}{\Delta_1 \Delta_2} \Gamma(P,q) \quad (57)$$

and, after reduction, the three-dimensional vertex function $\Gamma(\vec{q})$ is connected with the reduced BS function $\psi(\vec{q})$ by the simple formula

$$\psi(\vec{q}) = \frac{\Gamma(\vec{q})}{q^2 + m_q^2 - \frac{M^2}{4}}. \quad (58)$$

This formula is used for a calculation of the formfactor diagram. Straightforward calculation gives

$$F(Q_\perp^2) = \frac{2}{(2\pi)^3} \int \frac{d\vec{q}}{2\mu} \left(\frac{1}{4} - y^2\right) \psi(\vec{q}) \psi\left(\vec{q} + \frac{\vec{Q}_\perp}{2}\right) \quad (59)$$

(we neglect the longitudinal momentum transfer, $Q_+ \simeq 0$). From here one obtains the normalization condition:

$$\int \frac{d\vec{q}}{2\mu} \left(\frac{1}{4} - y^2\right) \psi^2(\vec{q}) = \frac{(2\pi)^3}{2}. \quad (60)$$

Now, using this equation and the relation $d\vec{q} = \mu d^2 q_\perp dy$, one has finally

$$\int d^2 q_\perp dy \left(\frac{1}{4} - y^2\right) \psi^2(q_\perp, y) = (2\pi)^3. \quad (61)$$

### D. Scattering amplitudes and nondiagonal transitions

Our next aim is to calculate amplitudes of elastic scattering of vector mesons on nucleons and, moreover, amplitudes of $VN \to V'N$ transitions between members of the $\rho$-meson family. The precise absolute values of these amplitudes are not so important rather we need only relative values, e.g., the ratio

$$F(\rho_0 N \to \rho_1 N) / F(\rho_0 N \to \rho_0 N).$$

Therefore, for the calculations the simplest model is used, namely, the well-known model of two gluon exchange [34, 35].

The nonperturbative effects are simulated in this model by introducing an effective gluon mass and an effective value of the quark-gluon coupling constant. The elastic amplitude for the meson-nucleon scattering is given, in this model, by the formula [47, 35]

$$F(s) = i\frac{16}{3} \alpha_s s \int \frac{d^2 q_\perp}{\left(q_\perp^2 + \mu_g^2\right)^2} \times \left[1 - F\left(4q_\perp^2\right)\right]\left[1 - F_N^{(Gauss)}\left(3q_\perp^2\right)\right], \quad (62)$$

where $\alpha_s (= g^2/4\pi)$ is the effective coupling constant, $\mu_g$ is the effective gluon mass which is a parameter of the model. $F$ and $F_N^{(Gauss)}$ are formfactors of the vector meson and the nucleon, respectively. For the latter the Gaussian approximation is used. For a calculation of the vector meson formfactor we use the expression (59). Firstly, perform the transformation to a transverse r-space using the convolution formula

$$\int d^2 q'_\perp dy \left(\frac{1}{4} - y^2\right) \psi(\vec{q}'_\perp, y) \psi(\vec{q}'_\perp + \vec{q}_\perp, y)$$
$$= \int d^2 r_\perp dy \left(\frac{1}{4} - y^2\right) e^{-i\vec{q}_\perp \vec{r}_\perp} \psi^2(r_\perp, y). \quad (63)$$

Substituting the expression for $F$ from eq. (59) in eq. (62) and using eq. (63) one obtains:

$$F(s) = i\frac{16}{3}\alpha_s s \int d^2 q_\perp \frac{V(q_\perp)}{\left(q_\perp^2 + \mu_g^2\right)^2}$$
$$\times \int d^2 r_\perp dy \left(1 - e^{-i\vec{q}_\perp \vec{r}_\perp}\right) \psi^2(\vec{r}_\perp, y), \quad (64)$$

$$V(q_\perp) \equiv 1 - F_N^{(Gauss)}\left(3q_\perp^2\right) = 1 - e^{-\frac{\langle r_N^2 \rangle}{2} q_\perp^2}. \quad (65)$$

Here, $\langle r_N^2 \rangle$ is the mean square radius of the nucleon. The expression for $F_N^{(Gauss)}$ in right hand side of eq. (65) is the Gaussian approximation for the nucleon formfactor [47].

The wave function in transverse r-space is obtained from $\psi(q_\perp, y)$ using the relation

$$\psi(r_\perp, y) = \frac{1}{2} \int dq_\perp^2 J_0(q_\perp r_\perp) \psi(q_\perp, y), \quad (66)$$

where $J_0(x)$ is the Bessel function.

For a calculation of amplitudes for nondiagonal transitions it is convenient to use a method of the diffraction-scattering eigenstates [48].

We assume that our amplitudes are provided by the elastic and single diffraction. The basis of diffraction scattering eigenstates is $|r_\perp, y\rangle$, and in this basis the scattering matrix is diagonal

$$\langle \vec{r}_\perp, y | \hat{F}(s) | \vec{r}'_\perp, y' \rangle = F_{\vec{r}_\perp}(s) \delta(\vec{r}_\perp - \vec{r}'_\perp) \delta(y - y'). \quad (67)$$

Vector meson states are expanded in a complete set of eigenstates:

$$|x\rangle = \int d^2 r_\perp dy \left(\frac{1}{4} - y^2\right)^{\frac{1}{2}} |\vec{r}_\perp, y\rangle \psi(\vec{r}_\perp, y) \quad (68)$$

The elastic scattering amplitude is now

$$F_{n,n}(s) = \langle V_n | \hat{F}(s) | V_n \rangle = \int d^2 r_\perp dy \left(\frac{1}{4} - y^2\right)$$
$$\times \psi^2(\vec{r}_\perp, y) F_{r_\perp}(s), \quad (69)$$

whereas, e.g., the amplitude of nondiagonal transition $V \to V'$ is

$$F_{n,n'}(s) = \langle V_n | \hat{F}(s) | V_{n'} \rangle = \int d^2 r_\perp dy \left(\frac{1}{4} - y^2\right)$$
$$\times \psi_{V'}(\vec{r}_\perp, y) \psi_V(\vec{r}_\perp, y) F_{r_\perp}(s). \quad (70)$$

The expression for the eigenamplitude $F_{\vec{r}_\perp}$ is obtained from comparison of eqs. (69) and (64) and is

$$F_{\vec{r}_\perp}(s) = i\frac{16}{3}\alpha_s s \int d^2 q_\perp \frac{V(q_\perp)}{\left(q_\perp^2 + \mu_g^2\right)^2}\left(1 - e^{-i\vec{q}_\perp \vec{r}_\perp}\right). \quad (71)$$

The formalism used here for a determination of vector meson-nucleon scattering amplitudes is very convenient: it gives a possibility to connect all these amplitudes (diagonal as well as nondiagonal ones) with each other because the expansion coefficients in eq. (68) are the vector meson wave functions which are derived using the same confining potential. We used above the approximation of spinless mesons and spinless quarks, so we will, naturally, regard the amplitudes $F_{n,n}$ and $F_{n,n'}$ in eqs. (69, 70) as amplitudes averaged over polarizations and denote them below by $\bar{F}_{n,n}$ (and $\bar{\sigma}_n$) and $\bar{F}_{n,n'}$. It is well known from photoproduction experiments (see, e.g., [54, 55]) that $\sigma_{\rho p}^T(s)$ and $\sigma_{\rho p}^L(s)$ are, in general, different and the ratio $\sigma^L(s)/\sigma^T(s)$ is, at small values of $s$ ($s \leq 10 \, GeV^2$), not smaller than ($0.25 \div 0.3$) [54]. Probably, this ratio grows with $s$, due to spin independence of diffraction scattering at large energies.

The relations based on VMD, as eqs. (32, 33), contain cross sections and nondiagonal amplitudes for transversely and longitudinally polarized vector mesons, $\sigma_n^{T,L}(s)$ and $F_{n,n'}^{T,L}(s)$. We will use the approximation

$$\sigma_n^T(s) \square \bar{\sigma}(s), F_{n,n'}^T(s) \square \bar{F}_{n,n'}^T(s)$$

and introduce the model parameter $\xi(s)$, which depends on the energy but does not depend on $n$, by the relation

$$\sigma_n^L(s) = \xi(s) \sigma_n^T(s), \quad F_{n,n'}^L(s) = \xi(s) F_{n,n'}^T(s).$$

Ending this Section one should note that we did not take into account rescattering of the eigenstates inside of the target and, correspondingly, did not consider unitarization problems. Due to this, there was no need to work with the general (nonforward) amplitude $F(s,t)$ and use the impact parameter space.

## IV. RESULTS OF CALCULATIONS

Structure functions of DIS in our two-component approach are sums of two parts, VMD (soft) component and pQCD (hard) component,

$$\sigma_{T,L}(Q^2, s) = \sigma_{T,L}^{soft}(Q^2, s) + \sigma_{T,L}^{hard}(Q^2, s).$$

The main formulas needed for calculations of the VMD (soft) contributions to the structure functions are eqs. (32, 33) in Section II. Photon-vector meson couplings $e^2/f_n^2$ are determined by the vector meson mass spectrum (eq. (54)) and by the quark-hadron duality relation (eq. (18)):

$$\frac{e^2}{f_n^2} = \frac{M_\rho^2}{M_n^2}\frac{e^2}{f_\rho^2}, \qquad \frac{f_\rho^2}{4\pi} = 2.25. \quad (72)$$

Cross sections $\sigma_n^T(s)$ and nondiagonal amplitudes $\frac{1}{s}F_{n,n'}^T(s)$ are calculated using eqs. (69, 70) for the amplitudes and eqs. (55, 66) for the vector meson wave functions. In the two-gluon-exchange approximation used for a derivation of eqs. (69,70) the $Vp$-scattering amplitudes divided on $s$ do not depend on the energy. Introducing now this dependence we assume that all vector mesons interact with the nucleon only nonperturbatively (in accord with VMD) and all amplitudes (including nondiagonal ones) have the same Regge-type energy dependence. For a normalization of this energy dependence one must use the necessary condition: the VMD prediction for a total photoproduction cross section for the real photon must agree with experimental data. Choosing the normalization point at $s = s_0$ one has the relations:

$$\sigma_n^T(0,s) = \sigma_n^T(0,s_0) f(s), \quad f(s_0) = 1, \quad (73)$$

$$\frac{1}{s} \mathrm{Im}\, F_{n,n'}^T(s) = \frac{1}{s} \mathrm{Im}\, F_{n,n'}^T(s_0) f(s). \quad (74)$$

It is meant here that amplitudes in right hand sides of these equations (which have argument $s_0$) are calculated using formulas of the two-gluon exchange, eqs. (69, 70, 71).

Now we can write down the expression for the photoproduction cross section (for the soft part of it),

$$\sigma_T^{soft}(0,s) = \sigma_{\gamma p}^{soft}(s) =$$

$$\left\{ \sum_n \frac{e^2}{f_n^2} \eta_T\left(M_n^2, k_{0\perp}^2\right) \sigma_n^T(s_0) \right.$$

$$\left. + \sum_{n \neq n'} \frac{e^2}{f_n f_{n'}} \eta_T\left(\max\left(M_n^2, M_{n'}^2\right), k_{0\perp}^2\right) \frac{1}{s_0} \mathrm{Im}\, F_{n,n'}^T(s_0) \right\} f(s). \quad (75)$$

If $s_0$ is low enough (we use, for concrete calculations, the value $\sqrt{s_0} = 8\,GeV$) a contribution of the hard component in $\sigma_{\gamma p}$ is small and the soft contribution can be safely normalized using the photoproduction data. It gives

$$\sigma_T^{(soft)}(0,s) = 114\,\mu b.$$

For the function $f(s)$ we use the Regge-type parameterization,

$$f(s) = \frac{c}{\sqrt{s}} + \left(\frac{s}{s_c}\right)^{0.06}, \quad (76)$$

where parameters $c$ and $s_c$ are not independent, due to the condition $f(s_0) = 1$.

For a description of the perturbative component we use the colour dipole model with the dipole cross section having a Regge-type $s$-dependence parameterized by the formula of FKS model [29]:

$$\sigma_{T,L}\left(Q^2, s\right) = \int dz d^2 r \left|\psi_\gamma^{T,L}(z,r,Q^2)\right|^2 \hat{\sigma}_{hard}(s,r), \quad (77)$$

$$\hat{\sigma}_{hard}(s,r) = \left(\alpha_2^H r^2 + \alpha_6^H r^6\right) e^{-\nu_H r} \left(r^2 s\right)^{\lambda_H}. \quad (78)$$

We used in calculations the following values of parameters:
$\alpha_2^H = 0.072$, $\alpha_6^H = 1.89$, $\nu_H = 3.27$, $\lambda_H = 0.44$.

Structure functions of DIS are defined by the expressions

$$F_2 = \frac{Q^2}{4\pi^2 \alpha}(\sigma_L + \sigma_T), \quad F_L = \frac{Q^2}{4\pi^2 \alpha}\sigma_L. \quad (79)$$

For a determination of the longitudinal cross section $\sigma_L^{soft}$ we use the eq. (33) and the parameter $\xi(s)$ defined above.

Our VMD formulas contain 4 main parameters: $\alpha_s, c, k_{\perp 0}$ and $\xi(s)$.

Parameter $\alpha_s$ enters the expression for the $Vp$-amplitude, eq. (62) and can be adjusted using the relation

$$\sigma_{\rho p} \simeq \frac{1}{2}\left(\sigma_{\pi^+ p} + \sigma_{\pi^- p}\right), \quad (80)$$

which is the prediction of additive quark model. The important parameter $k_{\perp 0}$ is determined (together with $s_c$) by a comparison of the model predictions with data for $\sigma_{\gamma p}(s)$ for the real photon and data for the structure function $F_2$ at small values of $Q^2$ (where a contribution of $\sigma_L(s)$ is small). At last, the $s$-dependence of $\xi(s)$ is adjusted using data for $F_2$ and, especially, for $F_L$ for relatively large $Q^2$, for which a contribution of the soft component is still essential.

The fitting of $\sigma_{\gamma p}, F_2, F_L$ gives, finally, the following values of model parameters:
$\alpha_s = 0.78$, $c = 1.15\,GeV$, $k_{\perp 0} = 0.385\,GeV$.

As for the function $\xi(s)$, our fitting gives the result:

$$\xi(s) = \begin{cases} 0.25, & s \leq 30\,GeV^2, \\ 0.17 \log s, & 30 \leq s \leq 7\,10^5\,GeV^2, \\ 1, & s \geq 7\,10^5\,GeV^2. \end{cases} \quad (81)$$

Main results of the calculations are shown on Figs. 5-9.

One of advantages of the VMD approach is the possibility to study the $Q^2 \to 0$ limit in formulas, i.e., to determine the $s$-dependence of the photoproduction cross section for the real photon (Fig. 5). The interesting question here is how large the hard contribution to $\sigma_{\gamma p}$ at $Q^2 = 0$ is, in the region of high energies. The calculation with our parameterizations, eqs. (76) and (78), gives

$$R = \sigma_T^{(hard)}(0,s) / \sigma_{\gamma p}^{total}(s) \approx 20\%$$

at $\sqrt{s} \simeq 300\,GeV$.

On Fig. 6 one can see that our two-component model is able to describe correctly the structure function $F_2(Q^2, x)$ in the region of low and medium $Q^2$ and in the $x$-interval $(10^{-4} \div 8 \cdot 10^{-2})$. On Fig. 7 we show the results for $F_L(Q^2, x)$, the structure function which is most sensitive to unknown function $\xi(s)$. It is seen that the agreement with data is acceptable at small $Q^2$ ($Q^2 \leq 5 \div 6\,GeV^2$).

On Figs. 8-9 we show results of our calculations in the region of very small x (up to $x \simeq 10^{-8} \div 10^{-9}$). The data are absent at $x \leq 10^{-6}$ for $Q^2 \leq 0.1\,GeV^2$ (Fig. 8) and at $x \leq 10^{-5}$ for $Q^2 \geq 1\, GeV^2$ (Fig. 9). To take into account phenomenologically the effects of gluon saturation we slightly modified the exponential term in the formula for $\hat{\sigma}_{hard}$ in eq. (78). Namely, we assume that $\nu_H$ slowly increases with a decrease of $x$, ($\nu_H = 3.27;\ 4;\ 5$) at, correspondingly, $x = 10^{-5}, 10^{-7}, 10^{-9}$. It is seen from left parts of Figs. 8, 9 that at smallest $Q^2$ the resulting $x$-dependence of $F_2$, at $x \leq 10^{-6}$, is almost coincides with the corresponding prediction of GBW-model [56[ whereas at $Q^2 \geq 1\, GeV^2$ the GBW saturation effects in the same region of $x$ are stronger. The accounting of the gluon saturation, although purely phenomenological, is necessary if one wants to estimate a relative contribution of the soft component at such small values of $x$. It is seen from right parts of Figs 8, 9 that this contribution is quite essential at $x \Box 10^{-8} \div 10^{-9}$ even at rather large values of $Q^2$.

In the present work we took into account in our VMD formulas the contribution from only one vector meson family (the $\rho$-family) although, of course, one must consider all vector mesons entering the current-field identity. It is clear, however, that the relative contribution of $\omega$ and $\varphi$ will be much smaller. SU(3)- symmetry predicts the ratios

$$\frac{1}{f_\rho^2} : \frac{1}{f_\omega^2} : \frac{1}{f_\varphi^2} = 9 : 1 : 2,$$

so, e.g., $\gamma\varphi$-coupling is weaker on a factor of 4.5 (according to leptonic width data, this factor is even larger, about 6). Besides, $\varphi$-meson is more heavy, and the contribution of its family will be relatively more cut by our $\eta$-factors. We plan to include $\varphi$ and $\omega$ mesons in our calculations in a future paper.

To illustrate the relative importance of diagonal and nondiagonal amplitudes, on the example of $\sigma_{\gamma p}^{soft}(s)$, we show in Table 1 placed in the Appendix the different items of the sum over $n, n'$ in $\sigma_T (0, s)$ of eq.(75).

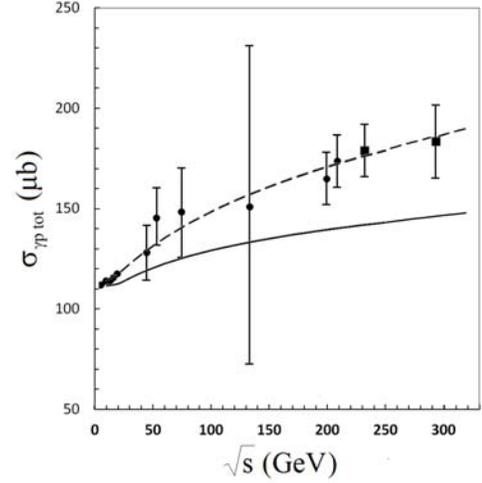

FIG. 5. The total cross section of photoabsorption for the real photon (the dashed line). The solid line is the soft contribution. The experimental data points are taken from [49-53]. The experimental points in the interval $\sqrt{s} = 40 \div 210\, GeV$ are taken from [49] (cosmic ray data).

On fig.1A of the Appendix we show how fast a value of the cross section is saturated with an increase of a number of vector mesons in the sum. It follows from the Table 1 that an addition of 9[th] meson changes the cross section on 0.44 %. The convergence of VMD sums at nonzero $Q^2$ is slightly worse but, at the same time, the relative contribution of the soft component decreases with $Q^2$. Our conclusion is that taking into account of 9 vector mesons in VMD sums is quite enough if $Q^2$ doesn't exceed 10-20 $GeV^2$: the corresponding error in a value of $F_2$ is smaller than (2-3)%.

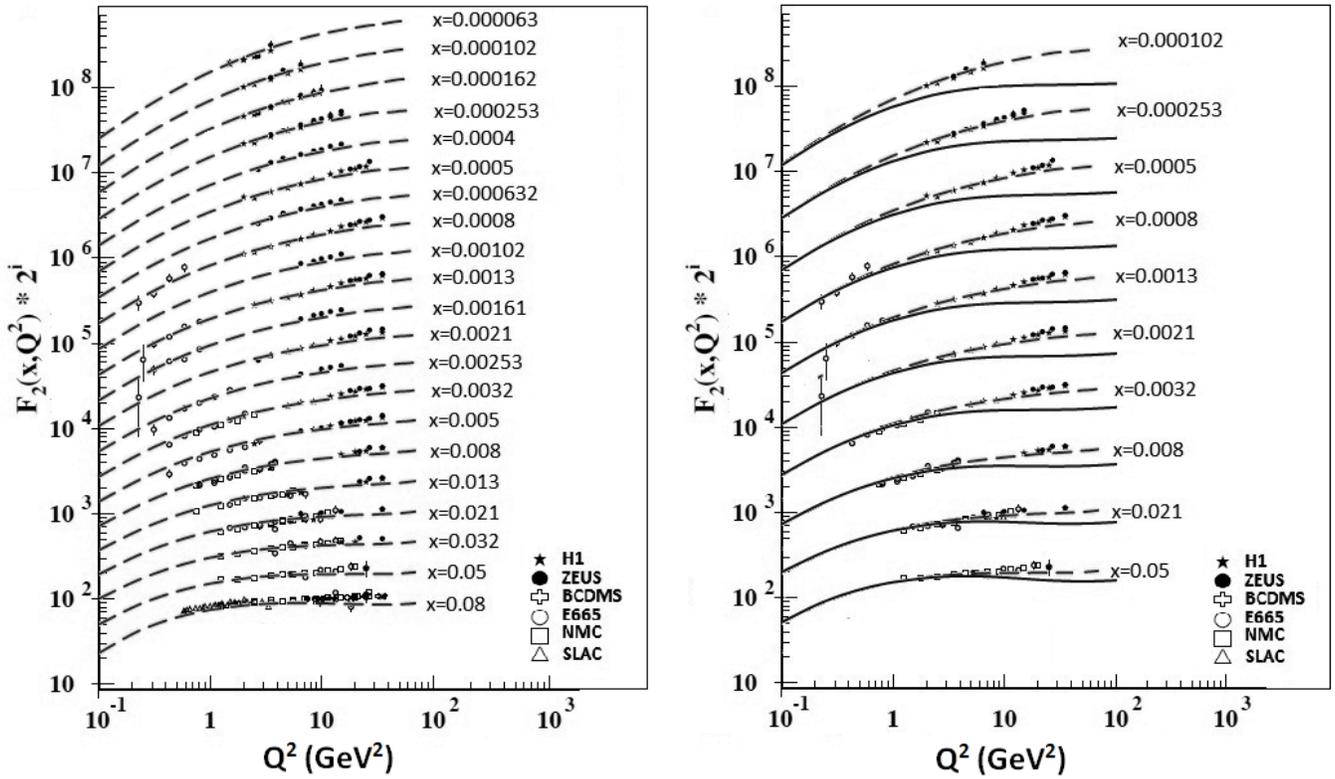

FIG. 6. The $Q^2$ dependence of the structure function $F_2$ for different values of $x$. On the left figure the data of each bin of fixed $x$ has been multiplied by $2^i$, where $i$ is a number of the bin, ranging from $i=8$ ($x=0.08$) to $i=28$ ($x=0.000063$). The experimental points are taken from [45]. The results of calculations (soft +hard) are shown by dashed curves. The right figure contains the same results but the data of x-bins are shown, together with theoretical curves, for the bins with odd numbers only (i=9,11…) and corresponding contributions of the soft component are shown separately, by solid curves.

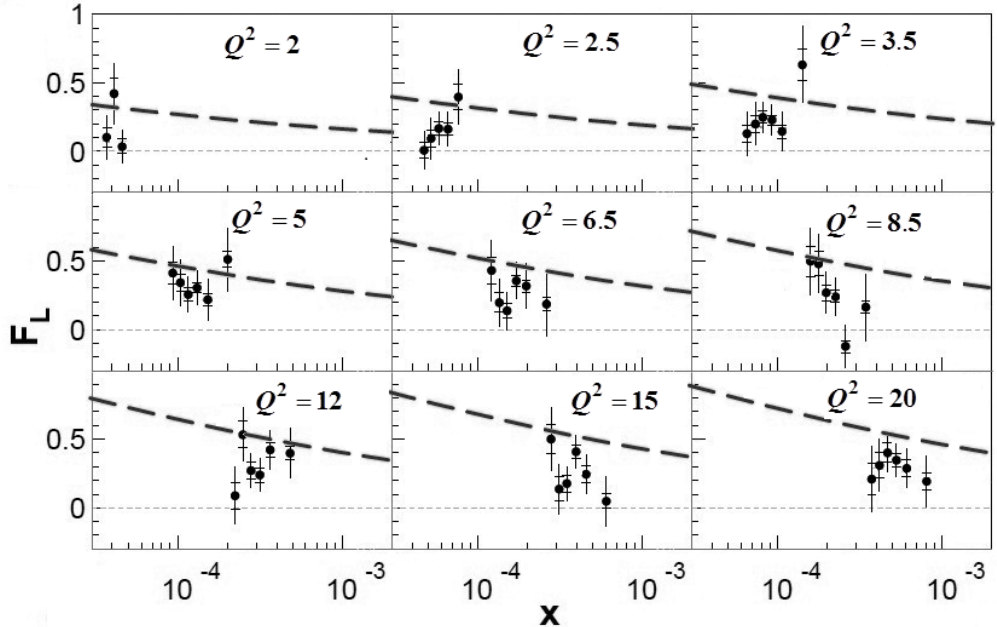

FIG. 7. The $x$ dependence of the structure function $F_L$ for different values of $Q^2$ $(GeV^2)$. The experimental points are taken from [57] (H1 Collaboration). The results of calculations (soft + hard) are shown by dashed curves.

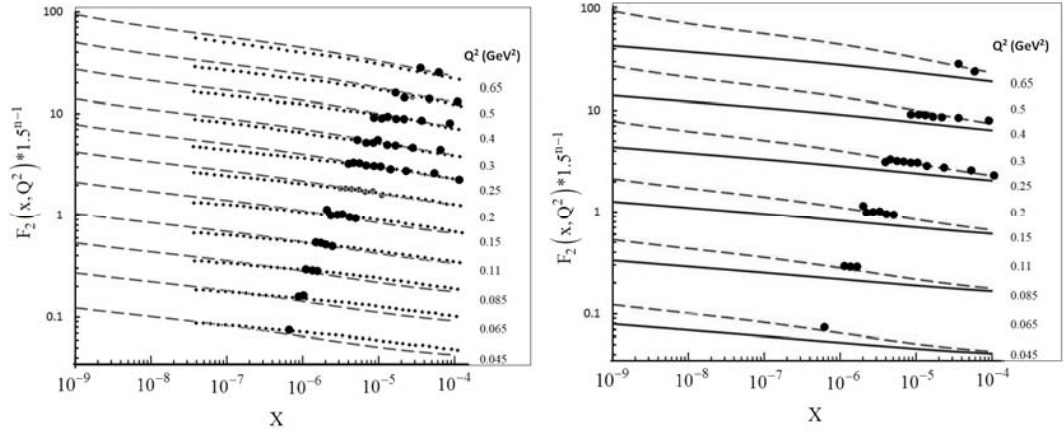

FIG. 8. The $x$ dependence of the structure function $F_2$ for small values of $Q^2$ in the region of very small $x$. The experimental points are taken from [58, 59]. The results of calculations (soft + hard) are shown by dashed curves. On the left figure the data are scaled by powers of 1.5, n = 1, 2, 3 … (from bottom to top). The dotted lines are the GBW predictions [56]. On the right figure the data are scaled by powers of 1.5, n = 1, 3, 5 … (from bottom to top), and the corresponding contributions of the soft component are shown separately, by solid lines.

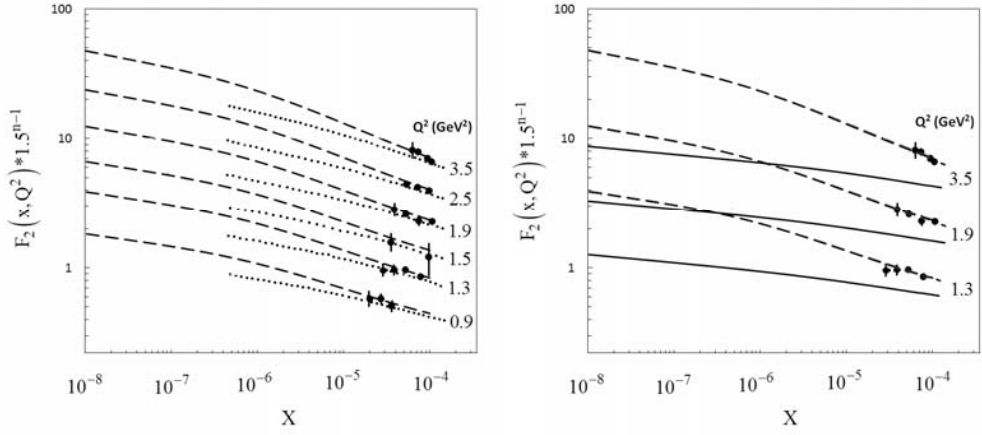

FIG. 9. The $x$ dependence of the structure function $F_2$ for medium values of $Q^2$ in the region of very small $x$. The experimental points are taken from [60]. The results of calculations (soft + hard) are shown by dashed curves. On the left figure the data are scaled by powers of 1.5, n=1,2,3…(from bottom to top). The dotted lines are the GBW predictions [56]. On the right figure the data are scaled by powers of 1.5, n=2,4,6…(from bottom to top), and the corresponding contributions of the soft component are shown separately, by solid curves.

## V. CONCLUSIONS

In the present paper we tried to show that the two-component description of the photon-nucleon inelastic scattering, with using VMD model as a nonperturbative component, is most natural and well-grounded theoretically. The main argument is quite simple: the vacuum fluctuations of real and weakly virtual photons are essentially hadronic (according to VMD hypothesis) and, therefore, just an use of VMD approach (which operates with hadrons rather than with quarks and gluons) is justified.

It is shown in the paper that the two-component model is successful in a description of experimental data for structure functions of DIS at small $x$ ($x<0.08$) and $Q^2<10\,GeV^2$. In the region of larger $Q^2$ the perturbative part begins to dominate and the whole description gradually becomes to be one-component. In the region of larger $x$, $x>0.1$, VMD concept doesn't work because the longitudinal size of photon's fluctuations becomes too small in comparison with the target's size.

The "deep diffraction" region of DIS, in which $Q^2<1\,GeV^2$, is the region where VMD component completely dominates if $x$ is larger than $10^{-4}$ (Fig.2) although in our model the aligned jet version of VMD suggested by Bjorken more than 40 years ago is exploited. This means that asymmetric configurations of $q\bar{q}$-pairs produced in photon's fluctuations are essential leading to a hadron-like interaction of the virtual photon with the nucleon. It is shown that even in the region of very small x, up to $x \sim 10^{-9}$, the soft component is noticeable, e.g., its contribution is about 40% at $x \sim 10^{-9}$, $Q^2=0.65\,GeV^2$ (Fig.8).

The predictions for the region of small x and $Q^2<1\,GeV^2$ are very important for applications to physics of very high energy cosmic rays. In particular, the so-called photonuclear energy losses of high energy cosmic ray muons in medium are determined just by the muon interactions in which a characteristic virtuality of the intermediate photon is small, $Q^2 \leq 1\,GeV^2$.

Finally, two remarks concerning, in particular, our plans on the future, are in order.

i) For a parameterization of a hard component of the structure functions we used the colour dipole model with a Regge-type s-dependence of the dipole cross section (eq. (78)). This parameterization had been suggested by authors of [29] as a contribution of the "hard pomeron". We did not modify the parameters adjusted in [29] except the region of very small $x$ ($x<10^{-5}$). Results of our paper show that the combination (VMD + FKS hard pomeron) gives rather good description of $F_2$ data at $Q^2$ smaller than $10-20\,GeV^2$. It means that our soft contribution (based, entirely, on VMD) and the soft pomeron component of FKS model give almost the same predictions for $F_2$ in low $Q^2$ region ($Q^2 \sim 1\,GeV^2$) where a hard contribution is small. As we see from Fig.6 our curves have a tendency to undershoot data. Surely, an agreement between our predictions and $F_2$ data at $Q^2$ larger than $10\,GeV^2$ can be improved by a better adjustment of the parameters entering eq.(78) with, probably, some modification of a form of this equation.

ii) Electromagnetic structure functions of hadrons in a region of large $Q^2$ are described quite well by approaches based on perturbative QCD taking into account, in particular, the gluon saturation effects. Global fits which use the pQCD predictions combined with some phenomenological ansatzes in small $Q^2$ region (see, e.g., the work of Albacete et al [27]) give the good description of $F_2$ over the full $Q^2$ range. Probably, it has a sense to try to use in our two-component approach, instead of the phenomenological Regge-type hard component, the pQCD-inspired component. It would essentially increase the range of $Q^2$ in which the two-component approach would give good predictions. Note, however, that the main motivation of our work was the description of structure functions just in the nonperturbative domain where a use of VMD concept is most natural.

**ACKNOWLEDGEMENTS**

Authors are grateful to Prof. A.I. Lebedev for useful discussions.

**APPENDIX**

TABLE 1. Each value in the table gives the contribution to $\sigma_{\gamma p}^{(soft)}(0,s)$ from one of transitions, $\gamma \to V_n \to V_{n'} \to \gamma$. $n$ is a number of the row, $n'$ is a number of the column, or vice versa. The total sum of all contributions is equal to $114\,\mu b$. $\sqrt{s}=8\,GeV$.

| $n \backslash n'$ | 0 | 1 | 2 | 3 | 4 | 5 | 6 | 7 | 8 |
|---|---|---|---|---|---|---|---|---|---|
| 0 | 70.94 | 10.16 | -2.61 | 0.78 | -0.44 | 0.18 | -0.13 | 0.07 | -0.04 |
| 1 | 10.16 | 10.27 | 1.94 | -0.68 | 0.21 | -0.14 | 0.05 | -0.04 | 0.02 |
| 2 | -2.61 | 1.94 | 4.48 | 0.86 | -0.36 | 0.12 | -0.09 | 0.03 | -0.03 |
| 3 | 0.78 | -0.68 | 0.86 | 2.56 | 0.49 | -0.23 | 0.08 | -0.06 | 0.02 |
| 4 | -0.44 | 0.21 | -0.36 | 0.49 | 1.66 | 0.31 | -0.16 | 0.06 | -0.05 |

| 5 | 0.18 | -0.14 | 0.12 | -0.23 | 0.31 | 1.17 | 0.21 | -0.12 | 0.04 |
| 6 | -0.13 | 0.05 | -0.09 | 0.08 | -0.16 | 0.21 | 0.88 | 0.15 | -0.09 |
| 7 | 0.07 | -0.04 | 0.03 | -0.06 | 0.06 | -0.12 | 0.15 | 0.68 | 0.11 |
| 8 | -0.04 | 0.02 | -0.03 | 0.02 | -0.05 | 0.04 | -0.09 | 0.11 | 0.54 |

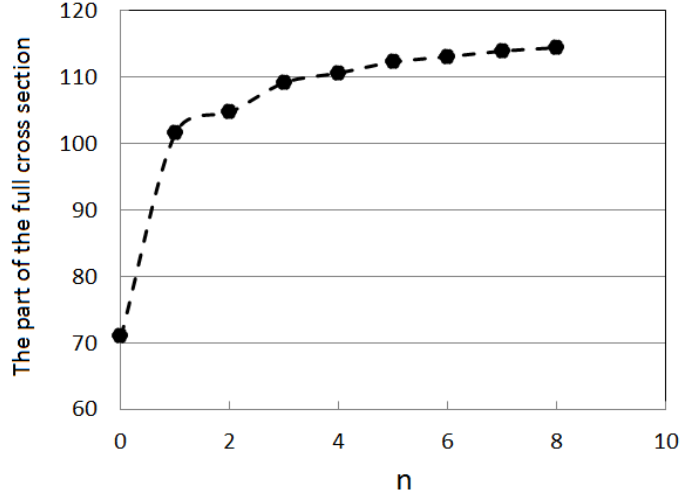

FIG. 1A. Horizontal axis: the number of vector mesons which is taken into account in a calculation of $\sigma_{\gamma p}^{(soft)}(0,s)$. Vertical axis: the corresponding part of the full cross section. The curve saturates on the value 114 $\mu b$ at $n=8$.

________________________________________________